# BETHE - SALPETER EQUATION – THE ORIGINS
Edwin E. Salpeter
October 2008

1. INTRODUCTION

The Bethe-Salpeter equation was first given at an American Physical Society meeting at the beginning of 1951 (Bethe-Salpeter 1951) and then published in more detail later in 1951 (Salpeter-Bethe 1951). The work was stimulated in part by two papers by Tamm and Dancoff, but then mainly by a series of seminal papers by Feynman and by Dyson on quantum electrodynamics, which started in 1949. The Bethe-Salpeter equation was an attempt to put bound states between two or more elementary particles on a fully covariant relativistic footing.

For a single electron, going from the non-relativistic Schrödinger equation (1926) to the fully covariant Dirac equation by 1929, went very quickly and there was an enormous amount of work during the 1930s on applications of non-relativistic quantum mechanics. It may then seem surprising for quantum electrodynamics and relativistic bound states to have waited 20 years or more. Of course, the second world war led to delays but these two topics in particular suffered from two great (and unrelated) difficulties. Quantum electrodynamics was held up by divergencies which were eventually helped by renormalization theory beginnings in 1947 (Sect. 2). For a fully relativistic treatment of bound states these divergencies were also a difficulty, but the main problem was (and to some extent still is) that each elementary particle had have its own time to go with its own spatial coordinates.

Although Bethe and I put in these separate times in our 1950 preparation for the 1951 paper, because we had to, we never felt comfortable with it. A famous theorist, when young and not yet famous, confided in Hans Bethe "I don't understand the Bethe-Salpeter equation". Hans' reply was: "Don't worry, neither do I really". This lack of understanding went for me in spades and it may seem surprising that we managed to use our equation to solve some explicit problems fairly quickly. However, these were problems of a special kind (Sect. 3).

2. DEVELOPMENTS BEFORE 1951

Dirac's relativistic quantum mechanics work in the 1920s started with a single electron in an atom, but the "hole theory" for dealing with the negative energy states soon raised multi-electron questions and also led to Fermi-Dirac quantum statistics. The various editions of Dirac's textbook (e.g. Dirac 1947) were eagerly read although not fully understood. At a 1947 Birmingham conference the opening reception boasted wall to wall Nobel Prize winners and Wolfgang Pauli, in his usual exuberance, insulted almost all of them but showed remarkable deference toward Dirac. When Richard Feynman published his seminal papers on Feynman graphs a couple of years later, he once remarked that an obscure long paragraph in Dirac's book foreshadowed his own work.

This Dirac-Feynman connection took 20 years, but Gregory Breit (Breit 1929) followed the Dirac 1928 work almost instantly with a paper on treating the interaction of two elementary particles relativistically. He did not use a separate time formalism and of course did not have the Feynman rules. He had an explicit wave equation, so he plus later collaborators (e.g. Breit and Brown 1948) could make explicit calculations. Unfortunately the lack of covariance made the equation somewhat ad hoc and only later work could show just which results were correct and which not. Nevertheless their concrete results helped later authors to develop their own work.

Three seemingly different versions of quantum electrodynamics were published by Tomonaga, Schwinger and Feynman in a series of papers starting 1948 for Tomonaga (in English, a year or two after the original), 1948 for Schwinger and 1949 for Feynman. These papers were all helped by the beginnings of renormalization theory, which in turn was sparked by the beautiful experimental work of Lamb and collaborators on the "Lambshift" (e.g. Lamb and Retherford 1947). These experiments showed level splittings which are not contained in the Dirac equation. Bethe (1947) both gave an approximate but adequate prescription for carrying out renormalization, which enabled one to deal with divergencies, and also a first theoretical calculation of the Lambshift itself.

Dyson's series of papers, starting with (Dyson 1949a) showed that the theories of Tomonaga, Schwinger and Feynman, although proposing different methods, were all equivalent according to basic quantum field theory. The Feynman formalism had the great practical advantage that terms could be characterized by the "Feynman graphs" or "Feynman diagrams" and then enumerated systematically. This was particularly useful for summing "ladder diagrams" where two elementary particles exchange many quanta but with no crossings of any quantum lines. We shall see that the Feynman formalism was particularly useful for formulating the Bethe-Salpeter equation (Sect. 3).

Tamm (1945) in Russia, with subsequent work in 1950, and Dancoff (1950)

in Illinois independently made some advances in treating bound states of pairs of elementary particles relativistically. Tamm's paper did not calculate explicit values for binding energy and was not well known in the West. Dancoff's paper got some binding energy values for a neutron and proton exchanging some kind of meson, but he did not consider his results close to the properties of a real deuteron. Unfortunately it was not clear how realistic the meson theory he used was and whether the discrepancy was due to the meson theory or due to inadequacies in Dancoff's method. Unfortunately, to my knowledge, Dancoff did not proceed further and he did not have the advantages of the Feynman formalism nor renormalization theory.

3. The 1951 PAPER AND EARLY APPLICATIONS

Bethe and I published our Salpeter and Bethe (1951) paper only a short while after Dancoff's, but we had the advantage of being at Cornell. The first renormalizaion theory paper was not only from Cornell but by Bethe himself (Bethe 1947) and Feynman was a professor at Cornell. The Feynman rules and Dyson analysis, with the clearcut separation into reducible and irreducible graphs, was particularly important for us: With a small coupling constant, the chance of a photon or meson being present at any one time is small, but with the infinite lifetime (or at least long) of a bound state there will be many steps in a ladder diagram (reducible), but with only a single (irreducible) exchange of a quantum.

The only explicit calculation of a binding-energy/coupling constant relation in our 1951 paper was also for two nucleons exchanging mesons, just like Dancoff's paper a short time earlier. As for Dancoff's paper, one did not know what the correct relation should be, but at least we showed that the calculation was fairly simple. Without explicit calculations, we also showed how more complex irreducible graphs could be treated within the Bethe-Salpeter equation formalism. Explicit calculations of various kinds in papers in the near future were very much on our mind.

There was only one paper, by Goldstein (1953), which tackled the most difficult case where the attractive coupling is so strong that the binding energy almost (or exactly) cancels the initial rest mass energy. This paper gave some interesting rules on properties of such bound states, if they existed at all. However, it was not possible at the time even to determine whether in fact such states existed.

In the two years after the (Salpeter/Bethe 1951) Bethe-Salpeter equation we published three papers which calculated small relativistic corrections to calculations which were already quite accurate. These corrections all involved the motion of a finite-mass nucleus in a hydrogen-like atom and the results could be compared with very accurate experimental values. There were two papers on the Mass Corrections for Fine Structure and Hyperfine Structure [Salpeter (1952); Salpeter and Newcomb (1952)] and soon after a paper on the Lambshift, Salpeter (1953). The corrections improved the already good agreement with experiments and made good propaganda for the Bethe-Salpeter equation.

The early 1950s provided another example of the West not appreciating work in Russia or Japan sufficiently. A paper by Hayashi and Munakata (1952) appeared in English just a short time after our 1951 B. S. equation, but with the Japanese precedents earlier. The work described there was extremely similar to the B. S. equation, but the equation continued to be called B.S., and not H.M., in the West. On the other hand, Chushiro Hayashi's classic work on stellar contraction down to the main sequence was given adequate publicity in the West.